\documentclass[twocolumn,aps,prl,beamer]{revtex4}%
\usepackage{amsmath}
\usepackage{amsfonts}
\usepackage{dsfont}
\usepackage{amssymb}
\usepackage{graphicx}
\usepackage{subfigure}
\usepackage{txfonts}%
\providecommand{\U}[1]{\protect\rule{.1in}{.1in}}

\begin{document}

\title{Topological Dirac semimetal phases in InSb/$\alpha$-Sn semiconductor superlattices}
\author{Jian-Peng Sun$^{1}$, Jia-ji Zhu$^{2}$, Dong Zhang$^{1}$\footnote{zhangdong@semi.ac.cn}, Kai Chang$^{1}$\footnote{kchang@semi.ac.cn}}
\affiliation{$^{1}$SKLSM, Institute of Semiconductors, Chinese Academy of Sciences,
P.O. Box 912, 100083, Beijing, China }
\affiliation{$^{2}$Institute for quantum information and spintronics, School of science, Chongqing University of Posts and Telecommunications, 400065, Chongqing, China }
\begin{abstract}
We demonstrate theoretically the coexistence of Dirac semimetal and topological insulator phases in InSb/$\alpha$-Sn conventional semiconductor superlattices, based on advanced first-principles calculations combined with low-energy $k\cdot p$ theory. By proper interfaces designing, a large interface polarization emerges
when the growth direction is chosen along {[}111{]}. Such an intrinsic
polarized electrostatic field reduces band gap largely and invert the band structure finally, leading to emerge of the topological Dirac semimetal phase with a pair of Dirac nodes appearing along the (111) crystallographic direction near the $\Gamma$ point. The surface states and Fermi arc are clearly observed in (100) projected surface. In addition, we also find a  two-dimensional topological insulator phase with large nontrivial band gap approaching 70 meV, which make it possible to observe the quantum spin Hall effect at room temperature. Our proposal
paves a way to realize topological nontrivial phases coexisted in conventional semiconductor superlattices by proper interface designing.
\end{abstract}


\maketitle

Band topology of solids has attracted intensive and broad interest in the past
decade due to the discovery of two-dimensional (2D) and three-dimensional (3D)
topological insulators.
Topological insulators posses a bulk gap and gapless boundary states protected
by the topology of bulk energy bands. These materials have been intensively
investigated both theoretically and experimentally in recent years because of
symmetry-protected dissipationless transport property \cite{PhysRevLett.95.226801,RevModPhys.82.3045,RevModPhys.83.1057,S. Murakami,A. Roth}.
The massless fermions emerge in
topological insulator edges/surfaces\cite{RevModPhys.83.1057,PhysRevLett.98.106803},
and topological Dirac and Weyl semimetals\cite{PhysRevLett.108.140405,PhysRevB.83.205101,PhysRevLett.107.127205,PhysRevX.5.011029}
. Among the distinct topological nontrivial phases, Dirac
semimetals are of particular interests, because the Dirac semimetals can be selectively
driven into other topological nontrivial phases such as topological
insulators and Weyl semimetals by specific symmetry breaking e.g., inversion
symmetry or time-reversal symmetry. Up
to date, various principles have been proposed to search and classify
Dirac semimetals protected by certain symmetry\cite{ncomms5898,PhysRevLett.115.126803}, and limited candidates
such as $A_3$Bi($A$=Na,K,Rb), $\text{Cd}_3\text{As}_2$ have been found\cite{PhysRevB.85.195320,PhysRevB.88.125427}.
 For deepening understanding of transitions
between distinct topological nontrivial phases and broadening the application horizon of the topological phases, it is highly desirable to construct topological insulators,
Weyl and Dirac semimetals by artificial designing in conventional semiconductors. The early attempts to design topological insulator phase in semiconductors were firstly proposed in the
GaN/InN/GaN\cite{PhysRevLett.109.186803} and later in
GaAs/Ge/GaAs\cite{PhysRevLett.111.156402} quantum wells (QWs) utilizing interface engineering.
Recently, a few experimental groups observe such huge interface electric field\cite{nphys1814,PhysRevB.88.125310} and the signature of topological phase\cite{1.4902916}. Recent theoretical works also prove this interface polarization induced topological insulator phases in other systems\cite{PhysRevLett.112.216803,nl5043769,adfm.201505357}.

\begin{figure}[tbph]
\centering \includegraphics[width=1\columnwidth]{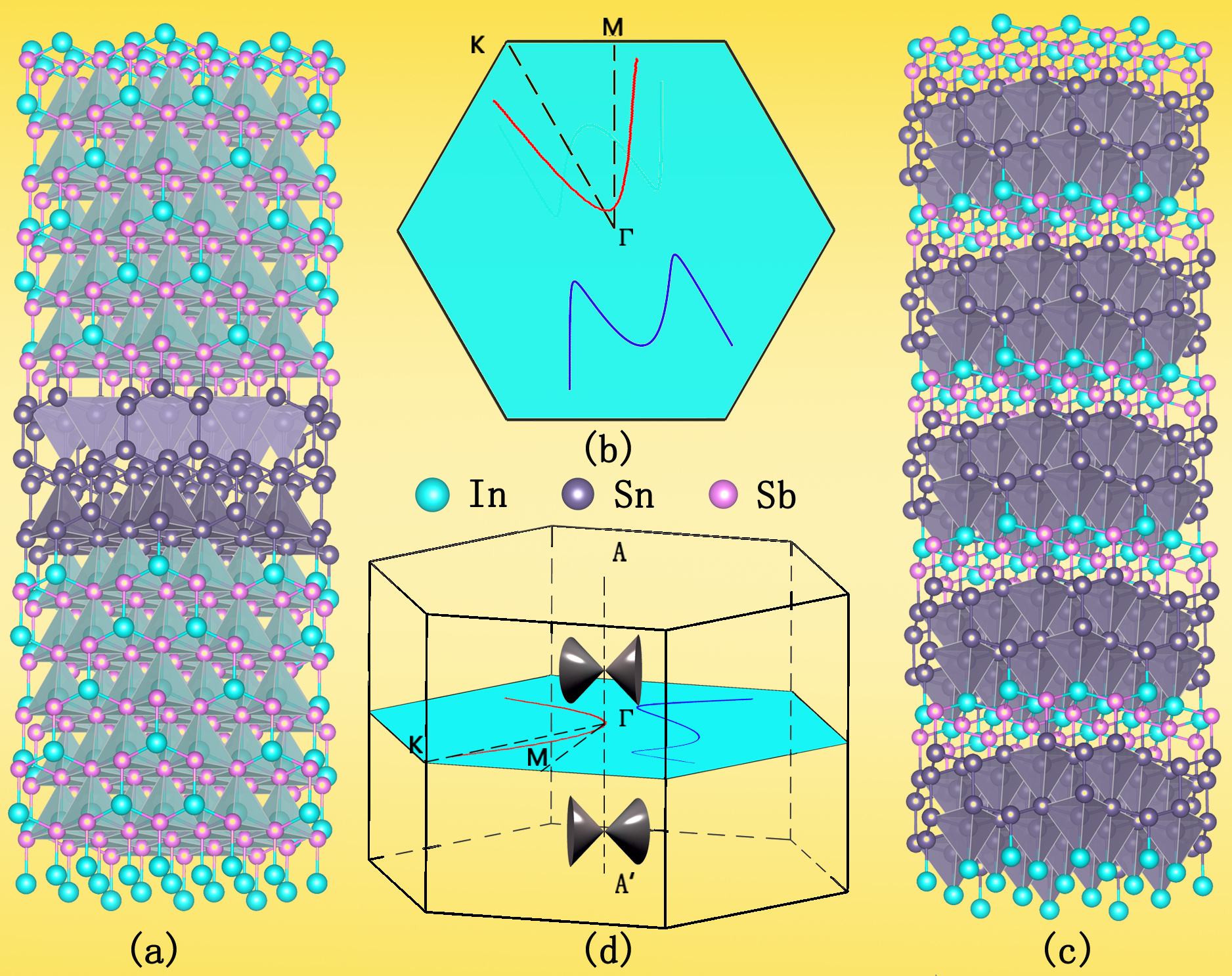} \caption{Schematics
of the InSb/$\alpha$-Sn semiconductor superlattices,
in which the indium, tin and antimony atoms are indicated by cyan,
grey and pink spheres, respectively. Panel (a) illustrates the atomic
configuration of a InSb/$\alpha$-Sn superlattice with thick InSb layers stacking along
the {[}111{]} polar orientation. Panel (c) : The same as (a), but for very thin InSb and $\alpha$-Sn layers, the corresponding energy dispersions and Dirac points near the $\Gamma$ points in the 3D and projected 2D Brillouin
zones are illustrated in (b) and (d).}
\label{model}
\end{figure}

In this work, we demonstrate theoretically that the Dirac semimetal phase can emerge in InSb/$\alpha$-Sn semiconductor
superlattices (SLs). $\alpha$-Sn is a
well-known semimetal\cite{Gapless-Semiconductors,PhysRevB.87.235307}
because the s-like $\Gamma_{6c}$ band is lower
than the p-like $\Gamma_{8v}$ band due to the relativistic mass-Darwin
effect.  $\alpha$-Sn can be tuned to be topological insulators by applying external
strain\cite{PhysRevLett.111.216401,PhysRevLett.111.157205}. However, ARPES and
 spin-resolved transport measurements reveal
that the Dirac-cones of the systems are buried below the Fermi surface
about 100 meV\cite{PhysRevLett.111.157205}. The zinc-blende InSb is a
semiconductor with a direct narrow bandgap located at $\Gamma$ point, and the lattice mismatch
between InSb (6.480{\AA}) and $\alpha$-Sn (6.489{\AA}) is about 0.14\%. Such a small lattice
mismatch can effectively decrease defects, vacancies, and dislocations
on the interfaces, and make InSb a perfect partner to resemble heterostructures
with $\alpha$-Sn. Molecular
beam epitaxy experiments reveal that InSb/$\alpha$-Sn/InSb
{[}111{]} heterostructures can be fabricated with sharp
interfaces and the interface polarization can be maintained therein\cite{PhysRevLett.72.2596},
which makes our proposal achievable experimentally accessible.

Fig. \ref{model} shows the superlattices consisted of
atomically-thin InSb and $\alpha$-Sn layers stacked periodically along the {[}111{]} direction of
the zinc-blende bulk materials. There is a huge intrinsic polarized electric field arising from opposite
charge accumulations at adjacent interfaces.
This electric field can reduce the bandgap significantly and generate a strong
Rashba spin-orbital coupling (SOC) effect. This field also depends sensitively on the thicknesses of $\alpha$-Sn and InSb layers.
For the thick InSb layer, the polarized electric field decreases rapidly and the electron wave functions in each $\alpha$-Sn layers are decoupled and are isolated completely, in analogy to a single QW case, as illustrated
as in Fig. \ref{model}(a) and (b). As the thickness of the InSb layers
decreases, the overlapping of adjacent $\alpha$-Sn layers is no longer
negligible and the energy dispersions along $k_z$ direction cannot be neglected, see Fig. \ref{model}(d).

In order to investigate the band structures of InSb/$\alpha$-Sn superlattices,
we adopted the advanced first-principle calculation\cite{PhysRev.136.B864} performed by the Vienna ab
initio Simulation Package (VASP)\cite{PhysRevB.54.11169} with the local
density approximation (LDA)\cite{PhysRev.140.A1133}. The electron-ion
interaction is described by pseudopotentials that are generated within
the projector-augmented wave (PAW) method\cite{PhysRevB.59.1758}. A plane wave basis set
with a kinetic energy cutoff of 560 eV for bulk calculations and the
superlattices is used. For bulk
Brillouin-zone integrations, a mesh of 11$\times$11$\times$11 $\Gamma$-centered
k points are used\cite{PhysRevB.13.5188}. For the thick SLs case, we use the
$\Gamma$-centered Monkhorst-Pack scheme to sample the Brillouin zone,
and the mesh of k-point sampling is 7$\times$7$\times$3. The convergence
criterion of the self-consistency process is set to be $10^{-6}$
eV. Since the accuracy of the band gap is particular importance
in our study, we applied Tran-Blaha method with the modified Becke-Johnson
(MBJ) semilocal exchange functional, named MBJLDA method\cite{PhysRevLett.102.226401}
to correct the underestimations of bandgap from traditional DFT-LDA. The
method has been proved successful and effective in many zinc-blende
compounds\cite{PhysRevB.82.205212,PhysRevB.80.035203} and in our cases
(see Supplementary Part I\cite{Supplement}), we find the equally
setting parameter CMBJ=1.200 for InSb and $\alpha$-Sn could produce
reliable band gaps for bulk materials and heterostures.
Moreover, to capture topological phase transitions in the heterostructures,
the SOC effects are taken into accounts in
all the calculations.

\begin{figure}[ptbh]
\centering \includegraphics[width=1\columnwidth]{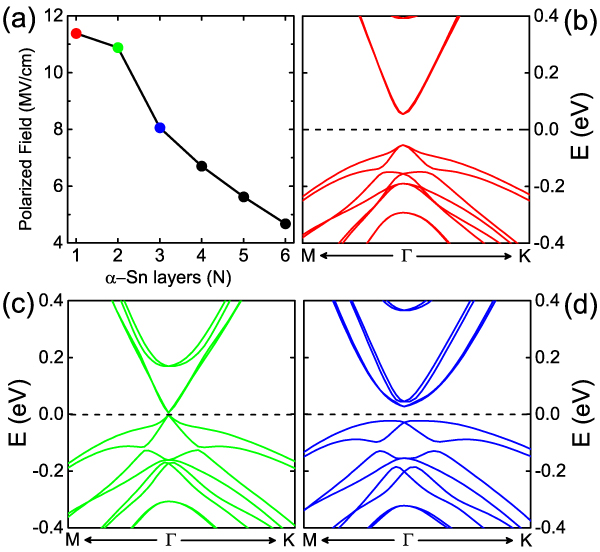} \caption{(color online)
The evolution of the interface polarized electric field strength and band structures of
SLs with the thickness of $\alpha$-Sn layers for thick InSb layers. (a) The polarized electric field strength as a function of the  number (N) of the $\alpha$-Sn atomic bilayer. The band structures of the SLs containing thick InSb layers but single $\alpha$-Sn bilayer (b) and two $\alpha$-Sn bilayers (c), three $\alpha$-Sn bilayers (d).  The panel (d) shows clearly the band inversion with nontrivial bandgap approaching 70 meV.}
\label{TI}
\end{figure}

First, we calculate the electronic structures of InSb/$\alpha$-Sn
SLs with thick InSb layers, to ensure isolation between electrons in adjacent layers,
modeled by supercells consisting of 12 atomic bilayers.
In order to understand the evolution of the electronic structures,
we shows the energy spectra for different $\alpha$-Sn bilayers (N=1,2,3) in Figs. \ref{TI} (b)-(d), respectively. Considering
the different atom bondings at the upper and lower interfaces
in the InSb/$\alpha$-Sn superlattice structures shown
in Fig. \ref{model}(a), Sn-Sb bonds at the
upper interface and Sn-In bonds at the lower interface, negative and positive
charges accumulate at upper and lower interfaces due to the charge transfer between Sn and Sb (In) atoms, respectively and create an intrinsic
polarized electric field crossing the $\alpha$-Sn
layers. From Fig. \ref{TI}(a), one can find out that the polarized
field can be as large as 11.2  MV/cm when there is only one bilayer
Sn sandwiched in two adjacent thick InSb layers and decreases rapidly to 4.8  MV/cm for six Sn bilayers due to the charge transfer in the adjacent interfaces. At the same time, the increasing of the thickness of
$\alpha$-Sn weakens
the quantum confinement effect dramatically, as a consequence, from Fig.
\ref{TI}(b), one can see that the quantum
confinement effect is so strong that the QW system possesses a normal band order and a trivial
band gap about 100 meV even under such a huge electric field (11.2 MV/cm). When the thickness of the Sn layers increases
to 3 bilayers, as illustrated in Fig. \ref{TI}(d), the polarized
electric field play a dominant role, the conduction and valence bands
are inverted with a nontrivial band gap as large as 70 meV, which means the quantum spin Hall effect in such systems can be observed at room temperature. This nontrivial gap is much larger than the nontrivial gaps in HgTe QWs (about 10meV) and InAs/GaSb QWs (about 5meV). This topological insulator phase transition occurs because of i) the decrease of band gap caused by the interface polarized electric field; and ii) the strong intrinsic SOC in $\alpha$-Sn
($\Delta_{SO}=0.7$ eV) layers and InSb ($\Delta_{SO}=0.8$ eV) layers. Fig. \ref{TI}(c) indicates the critical thickness of $\alpha$-Sn layer for the topological insulator phase transition. In the electronic
structures of a single $\alpha$-Sn bilayer sandwiched by thick InSb layers. A single four-fold degenerated Dirac point
emerges locating rightly at the Fermi energy, which can be viewed as a a 3D Dirac semimetal with anisotropic energy dispersion as an analogue to its 3D counterpart observed in HgCdTe semiconductor \cite{nphys2857}. This 2D Kane fermion system bridges the topological insulator phase and
trivial insulator phase. The 2D Kane Fermion can maintain
a 3D Dirac semimetal with anisotropic energy dispersion, e.g., quantum well case (see the on-line supplemental material, Part
II\cite{Supplement}).

\begin{figure}[ptbh]
\centering \includegraphics[width=1\columnwidth]{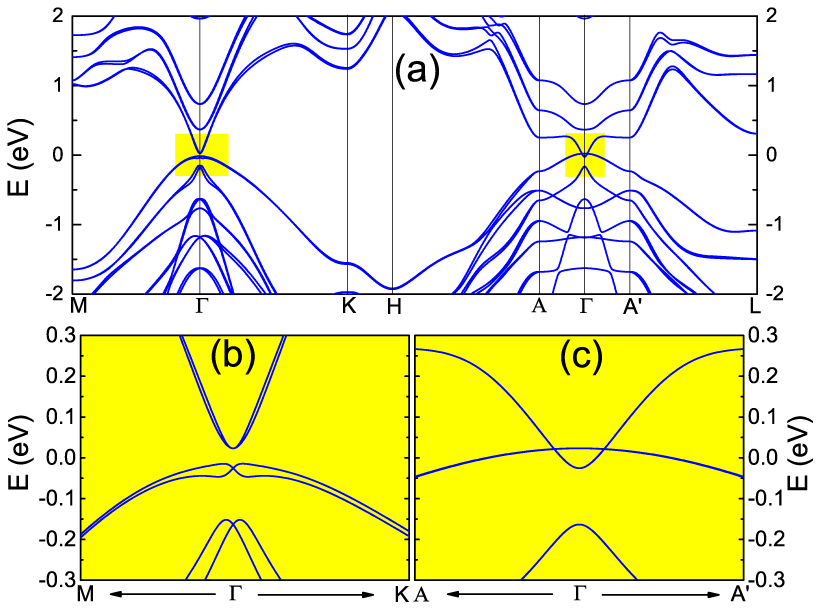} \caption{(color online)
(a) The band structures (including SOC) of InSb/$\alpha$-Sn
superlattice in the full Brilliouin zone (the upper panel),  the left-upper panels denotes the in-plane band structure along $M-\Gamma-K$ direction, while the right-upper panel describes the band structure along the growth direction of the SLs. The lower panels (b) and (c) amplify the highlighted regions in the upper panels (a). The highlighted
areas indicate nontrivial gap opening in vicinity of $\Gamma$
point (b), and in the panel (a) a pair of Dirac points emerging
along the $k_{z}$ axis (c).}
\label{DSM}
\end{figure}

\begin{figure}[ptbh]
\centering \includegraphics[width=1\columnwidth]{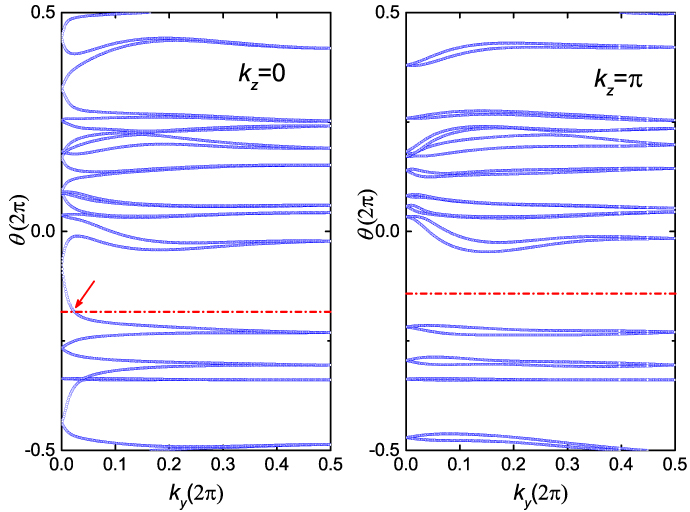} \caption{(color online)
 WCCs (Blue circles) evolution curves of InSb/$\alpha$-Sn superlattice in the 2D subsystems denoted by $k_z$ =0 (Left panel)and $k_z$ = $\pi$ (Right panel) of the unitcell. Possible reference lines are depicted in Red dashed lines. And an arrow indicates the crossing point of the WCC curves and the reference lines. }
\label{Z2}
\end{figure}

Next, we turn to search a topological Dirac semimetal phase in InSb/$\alpha$-Sn superlattices. We calculate the band structures of a series of short-period  InSb/$\alpha$-Sn superlattices. Considering
the shortest superlattice consisting of 3 atomic bilayers, in which
two bilayers of $\alpha$-Sn are sandwiched by one bilayer InSb to take
full advantage of polarized field and the quantum
confinement effect, the first-principles
calculation including the SOC reveals the SL is a semimetal. Fig. \ref{DSM}(a)
shows the band structures is gapless in the vicinity of the Fermi
energy in the full Brillouin zone.
The highlighted regions in Fig. \ref{DSM}(a)
are amplified in Figs. \ref{DSM} (b) and (c), respectively. Fig. \ref{DSM}(b) clearly depicts a nontrivial band gap
opening, i.e., the TI phase, just the same as in Fig. \ref{TI}(d), but with a smaller gap. More importantly, in Fig \ref{DSM}(c), a pair of Dirac points emerge at time-reversal points along the $k_z$ axis
in vicinity of $\Gamma$ point.
The two touching points possessing linear dispersions and keeping
unaffected in presence of SOC in companion with nontrivial $\mathds{Z}_2$ index indicate Dirac semimetal phase in the
superlattice. The topological indices of the InSb/$\alpha$-Sn superlattices are calculated by study of the evolution of the Wannier charge centers (WCCs) in $k_z$ =0 and $k_z$ = $\pi$ two-dimensional subsystems explicitly\cite{WCC1,WCC2}, see Fig.\ref{Z2}. The superlattice is topologically nontrivial since the number of transitions of WCC remains odd with respect to arbitrary reference lines.

Furthermore, the calculation of the Fermi surface and surface states
are performed to confirm the Dirac semimetal phase. We consider that a InSb/$\alpha$-Sn SL can be cleaved along a certain crystallographic plane,  forming a semi-infinite slab, a tight-binding Hamiltonian
was obtained with maximally localized Wannier function
(MLWF)\cite{PhysRevB.56.12847,PhysRevB.65.035109} derivated
from the first-principles calculations. The surface Green's function
of the semi-infinite slab was constructed from the tight-binding Hamiltonian
via iterative methods\cite{016,009}, whose imaginary part corresponds to the local density of states (LDOS) at the surface.

\begin{figure}[ptbh]
\centering \includegraphics[width=1\columnwidth]{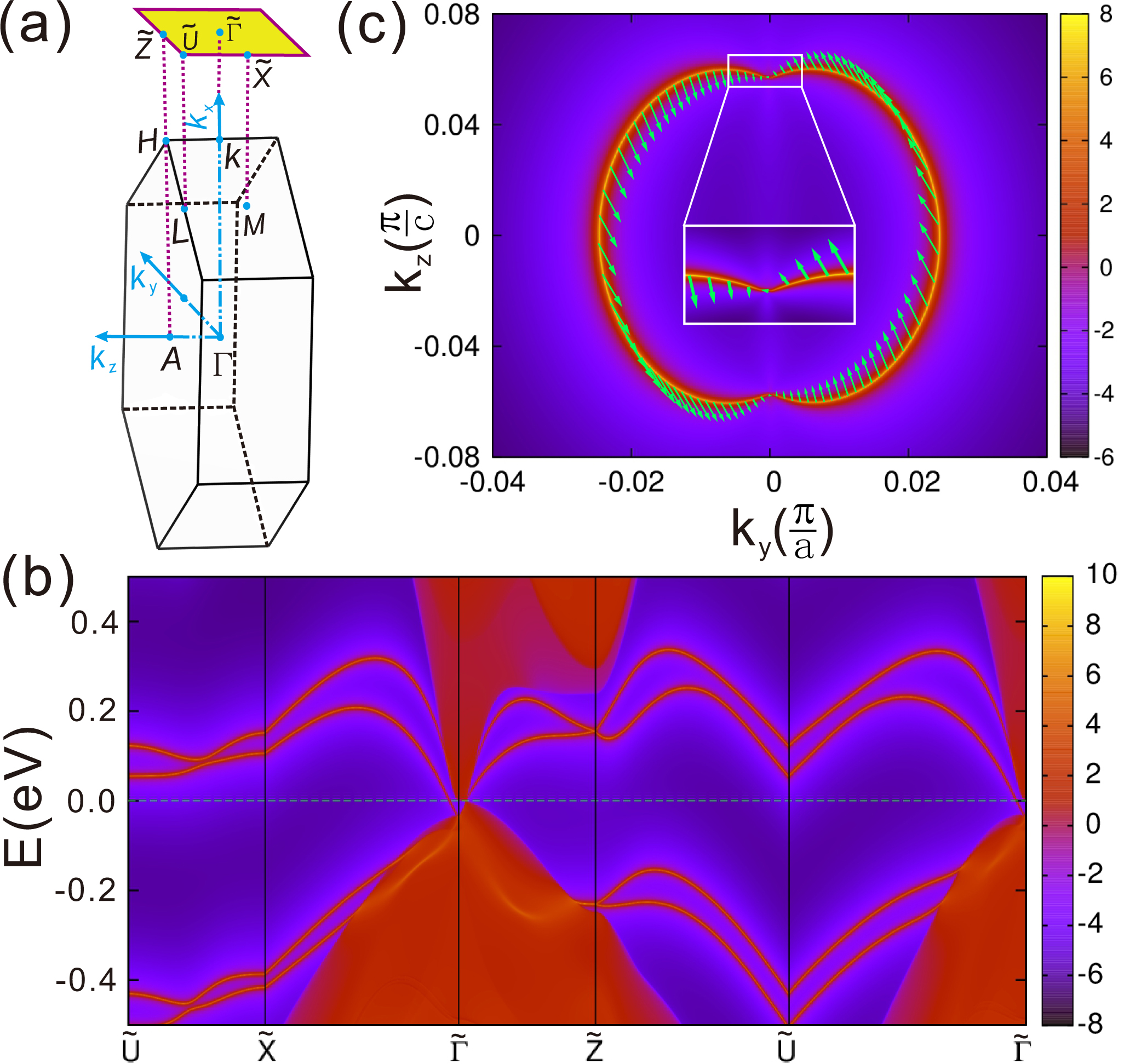} \caption{The Fermi loop
and surface state dispersion of the InSb/$\alpha$-Sn superlattice {[}100{]}
semi-infinite slab. (a) The folded Brillouin zone and its projection on (100) plane. (b) The LDOS
and energy dispersions of surface states and (c)The Fermi loop projected on the surface
(100) surface, in which the spin textures are denoted with the green arrows.}
\label{Fermi}
\end{figure}

If the InSb/$\alpha$-Sn superlattices are cleaved along (100) crystallographic surface, the energy dispersions of the bulk and surface states in the full Brillouin zone (see Fig. \ref{Fermi}(b)) are projected to a 2D Brillouin zone (see Fig. \ref{Fermi} (a)). One can clearly see a Fermi loop in Fig. \ref{Fermi}(c), which is composed of two half-circle Fermi arcs connected by two singular points which are corresponding to Dirac points in bulk band structure of the superlattice. The spin
texture is superposed on the Fermi loop by analysis of the local spin
density and suggests the chirality of the Fermi loop. In addition,
the dissipationless surface states are shown in Fig. \ref{Fermi}(b).
The dissipationless surface states in companion with the Fermi loop
are typical fingerprints of 3D topological Dirac semimetals.

\begin{figure}[pth]
\centering \includegraphics[width=1\columnwidth]{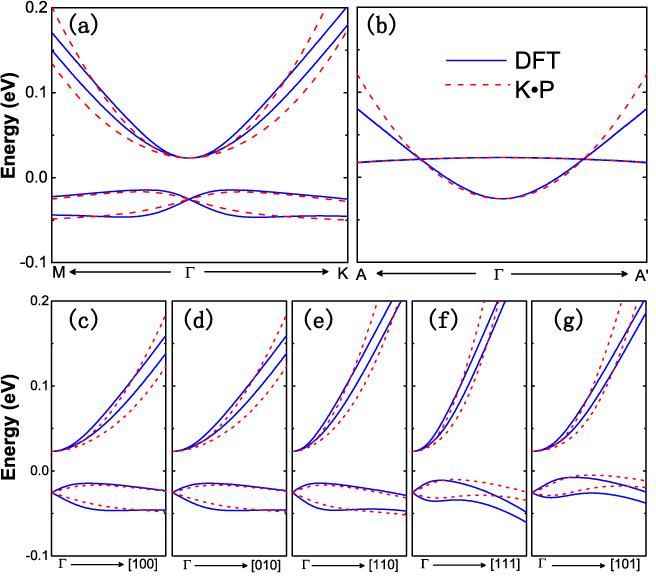} \caption{(color online)
Comparing between the results obtained from the first-principles calculation (the blue solid
lines) and effective $k\cdot p$ model (the red dashed lines) for in-pane (a) and out-of-plane directions. (a) Fitness in the
plane, (b) fitness along the $k_{z}$ vector and (c)-(g) fitness along arbitrary
reciprocal vectors. }
\label{fit}
\end{figure}

To capture the essential physics of the Dirac semimetal phase, we
deviated a low-energy effective $k\cdot p$ Hamiltonian. Since the first-principles
calculation unveil that, the low energy states mainly origin from
$s$ and $p$ orbital with two different angular momentum
$J=\frac{3}{2},\frac{1}{2}$, we chose the four bases set
of $\left\vert s,\frac{1}{2}\right\rangle ,\left\vert p,\frac{3}{2}\right\rangle ,
\left\vert s,-\frac{1}{2}\right\rangle ,\left\vert p,-\frac{3}{2}\right\rangle $.
By utilizing the theory of invariants\cite{PhysRevB.82.045122}, the effective $4\times4$
Hamiltonian around $\Gamma$ point reads
\[
H_{\Gamma}{\Large(\mathbf{k})=\epsilon(\mathbf{k})I}_{4\times4}+\begin{pmatrix}M(\mathbf{k})
& Ak_{+} & -iRk_{-} & B_{+}^{\ast}(\mathbf{k})\\
Ak_{-} & -M(\mathbf{k}) & B_{-}^{\ast}(\mathbf{k}) & 0\\
iRk_{+} & B_{-}(\mathbf{k}) & M(\mathbf{k}) & -Ak_{-}\\
B_{+}(\mathbf{k}) & 0 & -Ak_{+} & -M(\mathbf{k})
\end{pmatrix},
\]
where $\epsilon(\mathbf{k})=C_{0}+C_{1}k_{z}^{2}+C_{2}\left(k_{x}^{2}+k_{y}^{2}\right),
M(\mathbf{k})=M_{0}-M_{1}k_{z}^{2}-M_{2}\left(k_{x}^{2}+k_{y}^{2}\right),
k_{\pm}=k_{x}\pm ik_{y},B_{\pm}(\mathbf{k})=\left(\alpha k_{z}\pm\beta R\right)k_{+}^{2}$.
The Hamiltonian take very similar form as the model Hamiltonian for
topological insulators with new off-diagonal elements. The terms $B_{\pm}(\mathbf{k})$ containing
the high order contribution $\left(\alpha k_{z}\pm\beta R\right)k_{+}^{2}$ is
required by the threefold rotational symmetry, and the off-diagonal elements $iRk_{-}$ describes the SOC caused by the broken inversion symmetry.
Actually, one can get an effective Hamiltonian for describing InSb/$\alpha$-Sn/InSb
QWs by simply set $k_{z}$=0. And on the other hand, set $k_{x}$=$k_{y}$=0 and diagonalize
the effective Hamiltonian, one can get two Dirac points located at
$\vec{k}=\left(0,0,k_{z}^{D}=\pm\sqrt{\frac{M_{0}}{M_{1}}}\right)$.
For the specific superlattice in this work, we obtained the parameters
in the effective Hamiltonian by fitting energy spectrum of effective
model with that obtained from first-principles calculations, $C_{0}=-0.002\text{ eV}, C_{1}=78.331\text{ eV}\cdot\text{\r{A}}^{2}
,C_{2}=66.475\text{ eV}\cdot\text{\r{A}}^{2},
M_{0}=-0.024\text{ eV}, M_{1}=-84.906\text{ eV}\cdot\text{\r{A}}^{2},
M_{2}=-51.838\text{ eV}\cdot\text{\r{A}}^{2},
A=2.773\text{ eV}\cdot\text{\r{A}}, R=-1.419\text{ eV}\cdot\text{\r{A}}$,
where $M_{0},M_{1},M_{2}$ take minus signs to demonstrate the band
inversion. From Fig. \ref{fit} , one can find out that, the energy dispersions calculated
from the effective $k\cdot p$
Hamiltonian are in good agreement with the first-principles calculations in vicinity of the $\Gamma$ point.
Each fourfold degenerated Dirac point is created by band inversion,
and can be classified as topological Dirac semimetals created by band inversion in previous work\cite{ncomms5898}.
The topological Dirac semimetal phase can be widely maintained in InSb/$\alpha$-Sn
superlattices (see Supplementary Part III).

In summary, we studied systematically the electronic structures
of InSb/$\alpha$-Sn superlattices. The
first-principles calculations reveal the coexistence
of topological insulator phase and 3D topological
Dirac semimetal phases with a large nontrivial gap (70 meV) in the conventional semiconductor InSb/$\alpha$-Sn superlattice
by utilizing polarized interfaces therein, and the topological phases could
be tuned by well-developed semiconductor growth techniques.
Our proposal paves a way to design topological nontrivial phases in
semiconductor materials and provides a unified platform to observe topological
phase transitions at room temperature.

This work was supported by Grant No. 2015CB921503 from the
MOST of China and NSFC Grants No. 11504366, and No. 11434010.


\begin{thebibliography}{99}
\bibitem{PhysRevLett.95.226801} C. L. Kane and E. J. Mele,
\href{http://dx.doi.org/10.1103/PhysRevLett.95.226801}
{Phys. Rev. Lett. \textbf{95}, 226801 (2005)}.
\bibitem{RevModPhys.82.3045} M. Z. Hasan and C. L. Kane,
\href{http://dx.doi.org/10.1103/RevModPhys.82.3045}
{Rev. Mod. Phys. \textbf{82}, 3045 (2010)}.
\bibitem{RevModPhys.83.1057} X. L. Qi and S. C. Zhang,
\href{http://dx.doi.org/10.1103/RevModPhys.83.1057}
{Rev. Mod. Phys. \textbf{83}, 1057 (2011)}.
\bibitem{S. Murakami} S. Murakami, N. Nagaosa, and S. C. Zhang,
\href{http://dx.doi.org/10.1126/science.1087128}
{Science \textbf{301}, 1348 (2003)}.
\bibitem{A. Roth} A. Roth, C. Br\"{u}ne, H. Buhmann, L.W. Molenkamp,
J. Maciejko, X.-L. Qi, and S.-C. Zhang,
\href{http://dx.doi.org/10.1126/science.1174736}
{Science \textbf{325}, 294 (2009)}.
\bibitem{PhysRevLett.98.106803} L. Fu, C. L. Kane, and E. J. Mele,
\href{http://dx.doi.org/10.1103/PhysRevLett.98.106803}
{Phys. Rev. Lett \textbf{98}, 106803 (2007)}.
\bibitem{PhysRevLett.108.140405} S. M. Young, S. Zaheer, J. C. Y. Teo, C. L. Kane,
E. J. Mele, and A. M. Rappe,
\href{http://dx.doi.org/10.1103/PhysRevLett.108.140405}
{Phys. Rev. Lett. \textbf{108}, 140405 (2012)}.
\bibitem{PhysRevB.83.205101} X. Wan, A. M. Turner, A. Vishwanath, and S. Y. Savrasov,
\href{http://dx.doi.org/10.1103/PhysRevB.83.205101}
{Phys. Rev. B \textbf{83}, 205101 (2011)}.
\bibitem{PhysRevLett.107.127205} A. A. Burkov and L. Balents,
\href{http://dx.doi.org/10.1103/PhysRevLett.107.127205}
{Phys. Rev. Lett. \textbf{107}, 127205 (2011)}.
\bibitem{PhysRevX.5.011029} H. Weng, C. Fang, Z. Fang, B. A. Bernevig, and X. Dai,
\href{http://dx.doi.org/10.1103/PhysRevX.5.011029}
{Phys. Rev. X \textbf{5}, 011029 (2015)}.
\bibitem{ncomms5898} B. J. Yang and N. Nagaosa,
\href{http://dx.doi.org/10.1038/ncomms5898}
{Nature Communications \textbf{5}, 4898 (2014)}.
\bibitem{PhysRevLett.115.126803} S. M. Young and C. L. Kane,
\href{http://dx.doi.org/10.1103/PhysRevLett.115.126803}
{Phys. Rev. Lett. \textbf{115}, 126803 (2015)}.
\bibitem{PhysRevB.85.195320} Z. Wang, Y. Sun, X.-Q. Chen, C. Franchini, G. Xu, H.
Weng, X. Dai, and Z. Fang,
\href{http://dx.doi.org/10.1103/PhysRevB.85.195320}
{Phys. Rev. B \textbf{85}, 195320 (2012)}.
\bibitem{PhysRevB.88.125427} Z. Wang, H. Weng, Q. Wu, X. Dai, and Z. Fang,
\href{http://dx.doi.org/10.1103/PhysRevB.88.125427}
{Phys. Rev. B \textbf{88}, 125427 (2013)}.
\bibitem{PhysRevLett.109.186803} M. S. Miao, Q. Yan, C. G. Van de Walle, W. K. Lou,
L. L. Li, and K. Chang,
\href{http://dx.doi.org/10.1103/PhysRevLett.109.186803}
{Phys. Rev. Lett. \textbf{109}, 186803 (2012)}.
\bibitem{PhysRevLett.111.156402} D. Zhang, W. Lou, M. Miao, S.-C. Zhang, and K. Chang,
\href{http://dx.doi.org/10.1103/PhysRevLett.111.156402}
{Phys. Rev. Lett. \textbf{111}, 156402 (2013)}.
\bibitem{nphys1814} G. Singh-Bhalla, C. Bell, J. Ravichandran, W. Siemons, Y. Hikita, S. Salahuddin, A. F. Hebard, H. Y. Hwang, and R. Ramesh,
\href{http://dx.doi.org/10.1038/nphys1814}
{Nat. Phys. \textbf{7}, 80 (2011)}.
\bibitem{PhysRevB.88.125310} L. Zhou, E. Dimakis, R. Hathwar, T. Aoki, D. J. Smith, T. D. Moustakas, S. M. Goodnick, and M. R. McCartney,
\href{http://dx.doi.org/10.1103/PhysRevB.88.125310}
{Phys. Rev. B \textbf{88}, 125310 (2013)}.
\bibitem{1.4902916} W. Pan, E. Dimakis, G. T. Wang, T. D. Moustakas, and D. C. Tsui,
\href{http://dx.doi.org/10.1063/1.4902916}
{Appl. Phys. Lett. \textbf{105}, 213503 (2014)}.
\bibitem{PhysRevLett.112.216803} H. Zhang, Y. Xu, J. Wang, K. Chang, and S.-C. Zhang,
\href{http://dx.doi.org/10.1103/PhysRevLett.112.216803}
{Phys. Rev. Lett. \textbf{112}, 216803 (2014)}.
\bibitem{nl5043769} Q. Liu, X. Zhang, L. B. Abdalla, A. Fazzio, and A. Zunger,
\href{http://dx.doi.org/10.1021/nl5043769}
{Nano Lett. \textbf{15}, 1222 (2015)}.
\bibitem{adfm.201505357} Q. Liu, X. Zhang, L. B. Abdalla, A. Zunger,
\href{http://dx.doi.org/10.1002/adfm.201505357}
{Adv. Funct. Mater. \textbf{26}, 3259 (2016)}.
\bibitem{Gapless-Semiconductors} I. Tsidilkowski, \emph{Gapless
Semiconductors---A New Class of Materials} (Akademie, Berlin, 1988).
\bibitem{PhysRevB.87.235307} S. \"{K}ufner, J. Furthm\"{u}ller, L. Matthes,
M. Fitzner, and F. Bechstedt,
\href{http://dx.doi.org/10.1103/PhysRevB.87.235307}
{Phys. Rev. B \textbf{87}, 235307 (2013)}.
\bibitem{PhysRevLett.111.216401} Y. Ohtsubo, P. Le F\`{e}vre, F. Bertran, and
A. Taleb-Ibrahimi,
\href{http://dx.doi.org/10.1103/PhysRevLett.111.216401}
{Phys. Rev. Lett. \textbf{111}, 216401 (2013)}.
\bibitem{PhysRevLett.111.157205} A. Barfuss, L. Dudy, M. R. Scholz, H. Roth,
P. H\"{o}pfner, C. Blumenstein, G. Landolt, J. H. Dil, N. C. Plumb, M. Radovic,
A. Bostwick, E. Rotenberg, A. Fleszar, G. Bihlmayer, D. Wortmann, G. Li,
W. Hanke, R. Claessen, and J. Sch\"{a}fer,
\href{http://dx.doi.org/10.1103/PhysRevLett.111.157205}
{Phys. Rev. Lett. \textbf{111}, 157205 (2013)}.
\bibitem{PhysRevLett.72.2596} 31 H. Omi, H. Saito, and T. Osaka,
\href{http://dx.doi.org/10.1103/PhysRevLett.72.2596}
{Phys. Rev. Lett. \textbf{72}, 2596 (1994)}.
\bibitem{PhysRev.136.B864} P. Hohenberg and W. Kohn,
\href{http://dx.doi.org/10.1103/PhysRev.136.B864}
{Phys. Rev. B \textbf{136}, 864 (1964)}.
\bibitem{PhysRevB.54.11169} G. Kresse and J. Furthm\"{u}ller,
\href{http://dx.doi.org/10.1103/PhysRevB.54.11169}
{Phys. Rev. B \textbf{54}, 11169 (1996)}.
\bibitem{PhysRev.140.A1133} W. Kohn and L. J. Sham,
\href{http://dx.doi.org/10.1103/PhysRev.140.A1133}
{Phys. Rev. \textbf{140}, A1133 (1965)}.
\bibitem{PhysRevB.59.1758} G. Kresse and D. Joubert,
\href{http://dx.doi.org/10.1103/PhysRevB.59.1758}
{Phys. Rev. B \textbf{59}, 1758 (1999)}.
\bibitem{PhysRevB.13.5188} H. J. Monkhorst and J. D. Pack,
\href{http://dx.doi.org/10.1103/PhysRevB.13.5188}
{Phys. Rev. B \textbf{13}, 5188 (1976)}.
\bibitem{PhysRevLett.102.226401} F. Tran and P. Blaha,
\href{http://dx.doi.org/10.1103/PhysRevLett.102.226401}
{Phys. Rev. Lett \textbf{102}, 226401 (2009)}.
\bibitem{PhysRevB.82.205212} Y.-S. Kim, M. Marsman, G. Kresse, F. Tran, and P. Blaha,
\href{http://dx.doi.org/10.1103/PhysRevB.82.205212}
{Phys. Rev. B \textbf{82}, 205212 (2010)}.
\bibitem{PhysRevB.80.035203} Y.-S. Kim, K. Hummer, and G. Kresse,
\href{http://dx.doi.org/10.1103/PhysRevB.80.035203}
{Phys. Rev. B \textbf{80}, 035203 (2009)}.
\bibitem{Supplement} See Supplemental Material at,
\bibitem{nphys2857} M. Orlita, D. M. Basko, M. S. Zholudev, F. Teppe, W. Knap, V. I. Gavrilenko, N. N. Mikhailov, S. A. Dvoretskii, P. Neugebauer, and C. Faugeras,
\href{http://dx.doi.org/10.1038/nphys2857}
{Nat. Phys. \textbf{10}, 233 (2014)}.
\bibitem{WCC1} S. K\"ufner, L. Matthes and F. Bechstedt,
\href{http://dx.doi.org/10.1103/PhysRevB.93.045304}
{Phys. Rev. B \textbf{93}, 045304 (2016)}
\bibitem{WCC2} M. Taherinejad, K. F. Garrity, and D. Vanderbilt,
\href{http://dx.doi.org/10.1103/PhysRevB.89.115102}
{Phys. Rev. B \textbf{89}, 115102 (2014)}
\bibitem{PhysRevB.56.12847} N. Marzari and D. Vanderbilt,
\href{http://dx.doi.org/10.1103/PhysRevB.56.12847}
{Phys. Rev. B \textbf{56}, 12847 (1997)}.
\bibitem{PhysRevB.65.035109} I. Souza, N. Marzari, and D. Vanderbilt,
\href{http://dx.doi.org/10.1103/PhysRevB.65.035109}
{Phys. Rev. B \textbf{65}, 035109 (2001)}.
\bibitem{016} M. P. L. Sancho, J. M. L. Sancho, and J. Rubio,
\href{http://dx.doi.org/10.1088/0305-4608/14/5/016}
{Journal of Physics F: Metal Physics \textbf{14}, 1205 (1984)}.
\bibitem{009} M. P. L. Sancho, J. M. L. Sancho, J. M. L. Sancho, and J. Rubio,
\href{http://dx.doi.org/10.1088/0305-4608/15/4/009}
{Journal of Physics F: Metal Physics \textbf{15}, 851 (1985)}.
\bibitem{PhysRevB.82.045122} C. X. Liu, X. L. Qi, H. J. Zhang, X. Dai, Z. Fang, and S. C. Zhang,
\href{http://dx.doi.org/10.1103/PhysRevB.82.045122}
{Phys. Rev. B \textbf{82}, 045122 (2010)}.

\end{thebibliography}
\end{document}